# Better 3D Inspection with Structured Illumination Part I: Signal Formation and Precision

ZHENG YANG,[1] ALEXANDER KESSEL,[1] GERD HÄUSLER[1,*]

[1]*Institute of Optics, Information and Photonics, Friedrich-Alexander-University Erlangen-Nuremberg, Staudtstrasse 7/B2, 91058 Erlangen, Germany*
*Corresponding author: Gerd.Haeusler@physik.uni-erlangen.de*

**For quality control in the factory, 3D-metrology faces increasing demands for high precision and for more space-bandwidth-speed-product SBSP (number of 3D-points/sec). As a potential solution, we will discuss Structured-Illumination Microscopy (SIM). We distinguish optically smooth and rough surfaces and develop a theoretical model of the signal formation for both surface species. This model is exploited to investigate the physical limits of the precision and to give rules to optimize the sensor parameters for best precision or high speed. This knowledge can profitably be combined with fast scanning strategies, to maximize the SBSP, which will be discussed in paper part II.**

## 1. INTRODUCTION

This paper is motivated by the requirements for the 3D-inspection of electronic interconnects. There, the specifications are exemplary for high precision manufacturing:

rough and specular surfaces occur at the same sample,
the required precision is about 1 μm,
the required lateral resolution is in the 5 μm regime,
the depth of the measuring field is in the 1 mm range.

There are established sensor principles satisfying these specs: white light interferometry (WLI) and confocal microscopy (CM). However, there is an additional requirement: the optical 3D-sensor should deliver more than $10^8$ 3D-points/sec. No sensor displays all these features at the same time. In this paper we will explore an alternative method, Structured-Illumination Microscopy (SIM), for technical surfaces - which has the potential for very fast measurements. However, before discussing the options for high speed, the limits of the precision have to be investigated. We developed a model of the physical signal formation on smooth and rough surfaces. From this model, rules can be derived to optimize the sensor parameters, for the best precision or for high speed.

Optical 3D sensors have become standard for the inspection of precise industrial products. There is a wide spectrum of sensors at the market that display sufficient precision for many applications. The major issues today are a big lateral field, with at the same time high lateral resolution and, of course, "speed". Commonly the height of an object is measured in 3D-space. Assuming that the field size is $\Delta x \Delta y$ m² and the bandwidth is $1/(\delta x \delta y)$ m⁻², a single 3D height map will include $(\Delta x \Delta y)/(\delta x \delta y)$ 3D-points. The speed $H$/sec specifies that $H$ height maps can be acquired per second. So the "space-bandwidth-speed-product" SBSP=$H(\Delta x \Delta y)/(\delta x \delta y)$(1/sec) characterizes the number of acquired 3D-points per second. Of course, it is not just the SBSP that makes a sensor useful, the customer needs as well precision $\delta z$ and measuring range $\Delta z$. The product CC=SBSPlog$_2(1+\Delta z/\delta z)$ is the channel capacity of the sensor. CC gives a useful number (in bit/sec) for the performance of a sensor, in terms of speed and dynamical range [1]. It should be emphasized that all 3D-sensors need $E$ ($E \geq 1$) exposures to achieve one height map. In other words, the information efficiency $1/E$ is commonly less than unity [2,3].

We will now discuss candidates for the application mentioned above, to understand the origin of the SBSP limit or the CC limit.

Not discussed will be the paradigm of 3D-metrology: fringe projection. This method cannot, generally, measure specular surfaces. Moreover, it lacks depth of field for high lateral resolution requirements.

White light interferometry (WLI) [4,5] is nearly perfect in terms of precision and lateral resolution. It measures smooth and rough surfaces [6] with a field of view up to 200 mm [7]. As a unique feature, the precision is independent from the imaging aperture and from the working distance. So, narrow boreholes can be measured [8]. Smooth surfaces can be measured with nanometer precision while the precision on rough surfaces corresponds to the surface roughness [6]. But WLI is slow. The depth scan has to acquire the so called correlogram, which is modulated with a period of $\lambda/2 \sim 400$ nm. Even if the sampling theorem may be violated to a certain extent [9], the large number $E$ of necessary z-steps reduces the efficiency dramatically, and limits the speed of the $z$-scan to a few hundred micrometers per second which is a severe limitation for the achievable SBSP. Hybl et al. [2,3,10] described an idea to overcome this limitation. This method is not yet commercially implanted.

Confocal microscopy (CM) in its different implementations [11-15] can be much faster than WLI in terms of depth scan, because there is no scanning of a high-frequency carrier signal necessary. So CM might be a candidate for high speed measurement, however, CM displays a few drawbacks discussed below.

We will discuss a promising alternative: Structured-Illumination Microscopy (SIM): a pattern (commonly a sinusoidal fringe) is projected onto the object surface, through a micro objective (see Fig. 1). The surface with the scattered or reflected pattern is imaged via the same objective onto an image detector located in the conjugate image plane. While the object is scanned in depth (z-scan), the local fringe contrast $C(z)$ is measured, e.g., via phase shifting [16,17,22]. From the maximum of the contrast curve $C(z)$, the surface shape can be found. The major advantage of SIM, compared to WLI is: the scanning speed can be significantly improved, because, principally, only three sampling points over the width (FWHM) of the contrast curve are necessary to interpolate the maximum.

A few advantages of SIM over CM: the inherent partial spatial incoherence reduces speckle noise. SIM is technically less complex. Speed and precision can be flexibly balanced for each application. SIM can easily be scaled up for larger field. This might balance the better lateral resolution of CM, as for industrial 3D metrology the highest lateral resolution is commonly not that important compared to the biological applications.

Now the state of the art of SIM will be described: the idea to use the contrast of sinusoidal fringe images along the depth direction for the acquisition of 3D data, was originally published in 1988 [18]. The idea was improved in [16,17]: again, a sinusoidal grating is projected while the object is scanned along the optical axis, the fringe contrast is determined via phase-shifting. The method did not get that much attention, until Gustafsson et al. [19] implemented SIM for biological applications and demonstrated the two-fold improved lateral resolution.

In the meantime, different modifications of SIM are commercially available, for technical surfaces too. In [20], instead of a sinusoidal grating, multiple thin lines are projected and laterally shifted during the depth scan. The local height is calculated from the (maximal) gradient of the observed line intensity. Schwertner et al [21] avoid the lateral line scan by using a full-field sinusoidal grating. The contrast curve is calculated from two exposures (with phase 0 and $\pi$) at each z-position. In [23], a spatial convolution technique is applied to calculate the focus information.

In spite of being commercially used, to our best knowledge, there is no theoretical model describing the limits of precision at specular and rough surfaces. We want to understand and to give the rules which can help users to build "better SIM".

The discussion of "better SIM" will be split into two papers: In part I we will first deal with the theory of the signal generation and the physical limits of the precision. Part II is devoted to measurements with a big SBSP: we will explain how to avoid the crucial stop-and-go depth scan and at the same time achieve significantly improved information efficiency for speed improvement. Eventually a novel lateral scanning concept will be demonstrated that allows for high speed SIM with arbitrary field size, without sacrificing lateral and longitudinal resolution.

In this part I, we will model the signal generation for optically smooth and rough surfaces. Using this model, it is possible to analytically investigate the theoretically achievable limit of the precision. This enables to optimally adjust the system parameters for the best precision, and to understand the bottlenecks of SIM.

## 2. SIGNAL FORMATION

SIM is, basically, an active focus searching method. As illustrated in Fig. 1, a sinusoidal pattern is displayed by a FLCoS display and projected onto the object, with an incident light microscope. The focus plane of the observation is conjugate to the focus plane of the projection, so the maximum contrast of the grid image is observed if the object surface is in focus. While scanning the object in z-direction, we acquire $M$ phase shifted grating images at each z-position to calculate the contrast curve $C(z)$ according to Eq. (1) [24].

$$C(I_1, I_2 ... I_M) = \frac{2\sqrt{(\sum_{i=1}^{M} I_i \cos\frac{(i-1)2\pi}{M})^2 + (\sum_{i=1}^{M} I_i \sin\frac{(i-1)2\pi}{M})^2}}{\sum_{i=1}^{M} I_i}. \quad (1)$$

At least three images with three phase steps of 120° are necessary for a local contrast evaluation. In practice, four phase shifts are common to reduce artifacts from residual nonlinearity. With $M=4$, Eq. (1) degenerates to Eq. (1a):

$$C(I_1, I_2, I_3, I_4) = \frac{2\sqrt{(I_1 - I_3)^2 + (I_2 - I_4)^2}}{I_1 + I_2 + I_3 + I_4}. \quad \text{(1a)}$$

Eventually, the local "height" $z(x, y)$ in each pixel is calculated by localizing the peak of the contrast curve $C(z)$.

To investigate the limits of SIM, we first deal with the physics of the signal formation and derive the relation between the contrast curve and the sensor parameters.

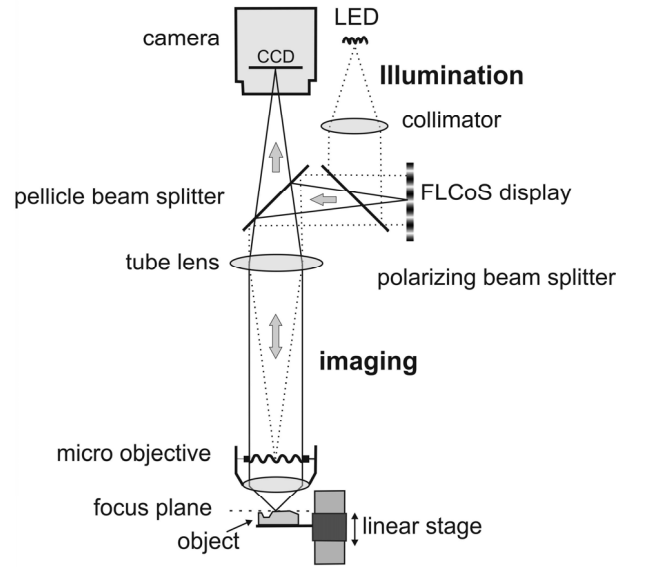

Fig. 1. SIM implemented in a bright-field microscope.

The model of the signal formation is derived by means of scalar diffraction theory and linear systems theory. In Fig. 2, the setup of Fig. 1 is unfolded. For easier understanding, the complete observation system is mirrored with respect to the object. Depending on the surface type (specular or scattering), the contrast has to be analyzed differently. For a specular surface, say, a planar mirror, a mirror image of the projected sinusoidal grating will be imaged onto the detector. If the object is out of focus by a distance $z_o$, the observed grating is defocused with effective defocusing $2z_o$. The effective pupil function of the total system (projection and observation) including defocusing distance $z_o$ is,

$$P_{\text{spec}}^{12}(\nu, \mu, z_o) = u_o \exp[i\pi\lambda 2z_o(\nu^2 + \mu^2)]. \quad (2)$$

where $\nu$ and $\mu$ are the spatial frequencies in $x$- and $y$- direction.

Note that the effective pupil function is just the product of the pupil functions for projection and observation, because reflection is a coherent process, for a specular object.

The incoherent OTF (optical transfer function) [25] is the normalized autocorrelation of the pupil function $P_{spec}^{12}$:

$$B_{spec}(\nu',\mu',z_o) = AC[P_{spec}^{12}(\nu,\mu,z_o)], \qquad (3)$$

$$OTF_{spec}(\nu',\mu',z_o) = \frac{B_{spec}(\nu',\mu',z_o)}{B_{spec}(0,0,0)}. \qquad (4)$$

The amplitude of the OTF displays the contrast function $C_{spec}$ on the detector (for a specularly reflecting surface).

$$C_{spec}(\nu',\mu',z_o) = |OTF_{spec}(\nu',\mu',z_o)|. \qquad (5)$$

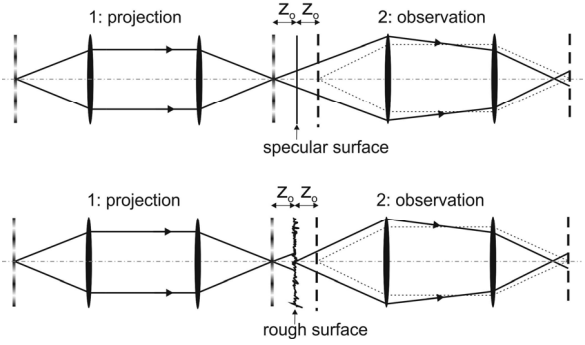

Fig. 2. Top: unfolded SIM setup for specular surface. Bottom: unfolded SIM setup for a scattering surface.

At the rough surface, as shown in Fig. 2 (bottom), the imaging process is different: a rough object, defocused by $z_o$, scatters a blurred grid image. Although intensity images observed by the camera suffer from coherent (speckle-) noise, we nevertheless may assume significant incoherence in the contrast map: As the illumination aperture is completely filled with the image of the light source, the diameter of the spatial coherence function at the object is small, according to the Van-Cittert-Zernike theorem. As a consequence, speckle contrast will be significantly reduced by defocusing during the depth scan. Furthermore, speckle noise is strongly correlated in phase shifted images and will be largely reduced after the contrast calculation (see section 3-B for details). So we may neglect speckle noise to a certain extent, in the modeling of SIM on rough surfaces.

The pupil function for the projection is,

$$P_{rough}^1(\nu,\mu,z_o) = u_o \exp[i\pi\lambda z_o(\nu^2 + \mu^2)]. \qquad (6)$$

where the defocusing term now is only $z_o$, instead of $2z_o$, compared to Eq. (2). The (incoherent) OTF of the projection is again given by the autocorrelation of the pupil function,

$$B_{rough}(\nu',\mu',z_o) = AC[P_{rough}^1(\nu,\mu,z_o)], \qquad (7)$$

$$OTF_{rough}(\nu',\mu',z_o) = \frac{B_{rough}(\nu',\mu',z_o)}{B_{rough}(0,0,0)}. \qquad (8)$$

The total OTF is square of the term in Eq. (8), and we get the contrast function $C_{rough}$:

$$C_{rough}(\nu',\mu',z_o) = |OTF_{rough}(\nu',\mu',z_o)|^2. \qquad (9)$$

The formalism above is principally textbook linear systems theory. Hopkins [26] derived an analytical solution for the defocused OTF. Stokseth gave a more easy-to-use numerical approximation [27] that we will further exploit. The simplified contrast curves on both surfaces types are:

$$C_{spec}(\bar{\nu},z_o) = \begin{cases} \left| 2(1 - 0.69\bar{\nu} + 0.0076\bar{\nu}^{-2} + 0.043\bar{\nu}^{-3}) \right. \\ \left. \times \left[ \frac{J_1(\alpha - 0.5\alpha)}{\alpha - 0.5\alpha} \right] \right| & |\bar{\nu}| < 2 \\ 0 & |\bar{\nu}| \geq 2 \end{cases}, \qquad (10)$$

$$C_{rough}(\bar{\nu},z_o) = C_{spec}(\bar{\nu}, \frac{z_o}{2})^2. \qquad (11)$$

with $\bar{\nu} = 2(\nu'^2 + \mu'^2)^{1/2} / \nu_{cutoff}$ and $\alpha = (2\pi/\lambda)\sin^2(u2z_o\bar{\nu})$, where $\bar{\nu}$ stands for the normalized spatial frequency $0 \leq |\bar{\nu}| \leq 2$, $J_1$ for the Bessel-function of first kind first order and $u$ for the full aperture angle in the object space.

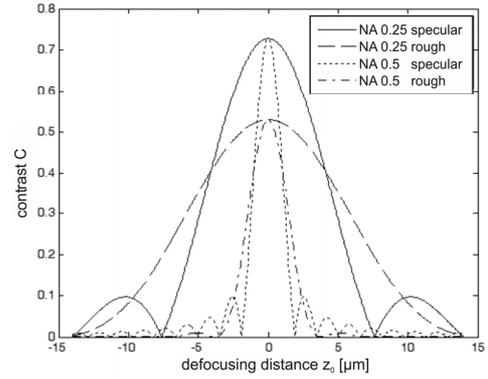

Fig. 3. Theoretical contrast curves $C(z_o)$ of a fringe pattern with $\nu'=0.2\nu_{cutoff}$, calculated by Eq. (10) and Eq. (11), for a specular and for a rough surface, each for two apertures NA=0.5 and NA=0.25.

Figure 3 displays contrast curves $C(z_o)$ for specular and rough surfaces. The contrast curves for a specular surface are more narrow (the ratio is 35/52), have a higher maximum and display some ringing. The width of the contrast curves goes with the inverse square of the aperture, as the Rayleigh depth of field.

According to Laboureux [28], the statistical uncertainty for the localization essentially depends on the curvature $\kappa_0$ at the maximum of $C(z_o)$ and on the noise. The curvature $\kappa_0$ can be mathematically calculated from the sensor parameters,

$$\kappa_0 \approx \frac{-3.921 C_{max} \sin^4 u\bar{\nu}(1-0.5\bar{\nu})^2}{\beta_1^2 \lambda^2}. \qquad (12)$$

with $C_{max}$ for the maximal contrast value of a contrast curve, $\beta_1=0.35$ for a specular object and $\beta_1=0.52$ for a rough object.
It is obvious that a higher and more narrow contrast curve (higher $\kappa_0$) is advantageous to obtain better localization. This will be discussed in the following section.

## 3. PRECISION
The precision is an important property in metrology. It represents the closeness between the results of repeated measurements under stipulated conditions [29]. The precision can be measured in the space domain. A paradigm example is taking a single measurement of a planar mirror: Assuming ergodicity, the precision is determined by calculating the standard deviation of the height map. For SIM, the precision is fundamentally limited by the localization uncertainty of the contrast maximum. Laboureux [28] investigated the ultimate lower limit of the registration precision of two one dimensional curves in presence of noise. We adapted the results for our task. As denoted in Eq. (13), the uncertainty of registration is dependent on the power spectral density $N_0$ of the (white) noise, the curvature $\kappa_0$ at the curve maximum and the full width of the data domain $\Delta T$ where the measurement values are selected for evaluation,

$$\sigma_z^2 = \frac{N_0}{1/12\kappa_0^2 \Delta T^3} = \frac{\sigma_C^2 \Delta s}{1/12\kappa_0^2 \Delta T^3}. \quad (13)$$

where $\sigma_C$ is the standard deviation of the contrast noise, $\Delta s$ is the sampling distance in depth direction and $N_0=\sigma_C^2 \Delta s$. The algorithms used in [28] and those for the height evaluation of SIM are similar: both methods use the least square method, so we can adapt Eq. (13) for SIM and derive a relation between the height uncertainty $\sigma_z$, the contrast noise level $\sigma_C$ and the sensor parameters. With Eq. (12) and Eq. (13), we obtain,

$$\sigma_z = \frac{\beta_2 \sigma_C \lambda}{\sqrt{N} C_{max} (\frac{\Delta T}{FWHM}) \sin^2 u \bar{\nu}(1-0.5\bar{\nu})}. \quad (14)$$

with $\beta_2$=0.31 for a specular object, $\beta_2$=0.46 for a rough object, $\sigma_C$ the standard deviation of contrast noise and $N$ the number of selected measurement values to be input to evaluation algorithm.

Equation (14) expresses quantitatively the precision of SIM for a given noise $\sigma_C$ of the contrast function $C(z_o)$ and given sensor parameters.

We move on to understand the fundamental origins of the precision. For this purpose all possible noise sources are analyzed and the dominant source of noise is identified.

**A. Specular surfaces**

At specular surfaces, the essential sources of noise are:

photon noise,
electronic (camera) noise,
statistical error of the translation stage,
nonlinearity of the sinusoidal pattern.

The first source is fundamental, while the other noise sources are caused by technology. Assuming that the technological errors are avoidable, the best achievable precision on specular surfaces is fundamentally limited by photon noise.
The photon noise in all phase shifted intensity images $I_i$, from which the contrast is calculated, is random and uncorrelated. The standard deviation $\sigma_C$ of the contrast noise can be expressed by Eq. (15) (we do not show the derivation and refer to the soon coming PhD thesis [30]). According to Eq. (15), $\sigma_C^2$ is inversely proportional to the squared SNR of the photon noise and the number $M$ of phase shifts. Furthermore, it depends on the contrast $C$ of the observed sinusoidal fringes.

$$\sigma_C^2 = (\frac{\partial C}{\partial I_1})^2 \sigma_{I_1}^2 + (\frac{\partial C}{\partial I_2})^2 \sigma_{I_2}^2 + ...(\frac{\partial C}{\partial I_M})^2 \sigma_{I_M}^2 \approx \frac{4}{SNR_{shot}^2 M}(1-\frac{C^2}{2}). \quad (15)$$

We combine Eq. (15) and Eq. (14) and get the ultimate relation for the dependence of the height precision from the sensor parameters. We use only the contrast values within the FWHM range (in other words: $\Delta T$/FWHM = 1).

$$\sigma_z = 0.62 \frac{1}{\sqrt{N}} \frac{1}{\sqrt{M}} \frac{1}{SNR_{shot}} \frac{\lambda}{\sin^2 u} \frac{\sqrt{\frac{1}{C_{max}^2}-0.5}}{\bar{\nu}(1-0.5\bar{\nu})}. \quad (16)$$

In Figure 4, the theoretical precision $\sigma_z$ calculated from Eq. (16) is compared to experimental results achieved with various micro objectives. The precision is depicted as a function of the chosen normalized fringe frequency $\bar{\nu}$. Curve 3 is calculated from Eq. (16) for a diffraction limited objective. In this case, the maximum contrast $C_{max}$ in Eq. (16) is the diffraction limited MTF from Eq. (5), for defocusing distance $z_o$=0 and fringe frequency $\bar{\nu}$. Curve 2 is again calculated from Eq. (16), however the (non diffraction limited) real maximum contrast $C_{max}$ was measured and used for the calculation. Eventually, curve 1 displays the experimentally determined precision, taken from the measurement of a planar mirror.

As the experimental setup displays additional error sources besides the photon noise, we are not surprised that the experiment (curve 1) shows a bigger uncertainty than curve 2. However, the trend of the results displays a considerable consistence between curve 1 and curve 2.

Figure 4(a) displays the results for a numerical aperture NA=.15. The precision (for curve 1 and 2) displays a soft minimum for frequencies around 10% of the cutoff frequency, whereas the theoretically optimal frequency lies at 25%. This is reasonable, as for the non diffraction limited objective, the minimum precision is achieved for a lower frequency than for the diffraction limited lens. It is useful to know that the minimum is not sharp. So there is no fine tuning of the projected fringe frequency necessary. It is further noteworthy that the ultimate limit of the precision for the ideal lens is about 2x better than for the real lens.

As illustrated in Fig. 4(a), a theoretical height uncertainty of around 200 nm can be achieved with a 5x/0.15 micro objective and standard video equipment. The precision can be extremely improved, down to a few nanometers, with higher NA, see, e.g., Fig. 4(d). Further improvement is possible by more samples for the depth scan and more phase shifts.

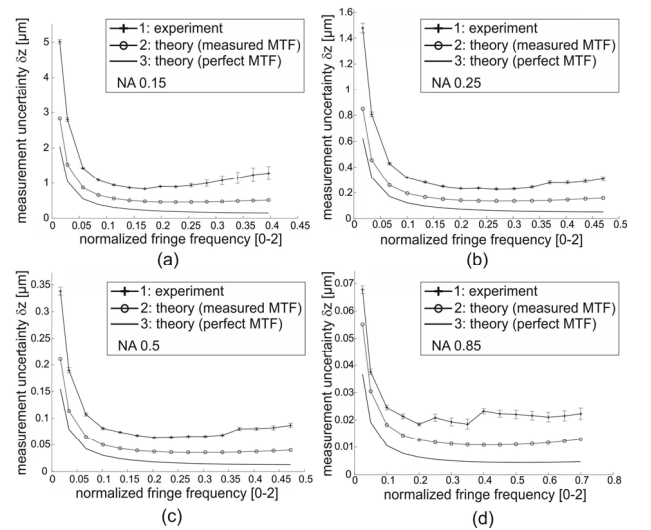

Fig. 4. Comparison of the theory of photon-noise limited measurement uncertainty and experimental results, for specular objects. The experiments are performed on a planar mirror, with 5x/0.15- (a), 10x/0.25- (b), 20x/0.5- (c) and 50x/0.85- (d) micro objective, where $\lambda$=0.5 µm, $N$=3, $M$=4. A signal-to-noise ration SNR$_{shot}$≈100 is assumed.

The results of Eq. (16) can easily be interpreted in a physical manner:

We calculate the contrast from $N$ samples within the contrast function, and from $M$ phase shifts. So we profit from some averaging and get the factors $1/N^{1/2}$ and $1/M^{1/2}$.

The term $\lambda/\sin^2 u$ displays the Rayleigh depth of field. The term $(1/C^2-0.5)^{1/2}/[\bar{\nu}(1-0.5\bar{\nu})]$ shows a global minimum around 1/4th of the cutoff frequency, for diffraction-limited optics. This term is softly varying around the optimal frequency, so we can choose the fringe frequency within a comparably wide range.

### B. Rough surfaces

If the object surface displays a roughness of larger than $\lambda/4$ within the width of the PSF, and for spatial coherence, speckle noise will be dominant [31], rather than photon noise. As illumination and observation use the same lens, the illumination aperture cannot be larger than the observation aperture. Hence, there is significant spatial coherence, even with a large light source.

Speckle noise in each phase-shifted image $I_i$ is highly correlated. This has to be taken into account for a quantitative evaluation of the precision. Keeping the analysis as much comprehensive as possible, we rewrite Eq. (1) as Eq. (17) where the contrast can be simplified as a function of $A$ and $B$,

$$C(I_1, I_2 ... I_M) = \frac{A-B}{A+B}, \quad (17)$$

with

$$A = \frac{1}{M}(\sum_{i=1}^{M} I_i + 2\sqrt{(\sum_{i=1}^{M} I_i \cos\frac{(i-1)2\pi}{M})^2 + (\sum_{i=1}^{M} I_i \sin\frac{(i-1)2\pi}{M})^2}) \quad (18)$$
$$= I_{mean} + CI_{mean},$$

and

$$B = \frac{1}{M}(\sum_{i=1}^{M} I_i - 2\sqrt{(\sum_{i=1}^{M} I_i \cos\frac{(i-1)2\pi}{M})^2 + (\sum_{i=1}^{M} I_i \sin\frac{(i-1)2\pi}{M})^2}) \quad (19)$$
$$= I_{mean} - CI_{mean}.$$

In Eq. (18) and Eq (19), $A$ and $B$ are numerically equivalent to the maximal and minimal intensities of the sinusoidal fringes.

We know from the experiments shown in Fig. 5(a) that speckle noise in all phase shifted images is highly correlated. So we assume in a first approach that it acts like spatial multiplicative noise $\rho(x,y)$, with a standard deviation $\sigma_\rho \in [0, 1]$. The intensity error $\Delta I_i$ can be expressed by Eq. (20),

$$\Delta I_i = I_i \rho(x,y). \quad (20)$$

From Eq. (20), the random error $\Delta A$ and $\Delta B$ in $A$ and $B$ can be approximated as in Eq. (21),

$$\Delta A = A\rho(x,y), \Delta B = B\rho(x,y). \quad (21)$$

Defining the speckle contrast $S_C = \sigma_\rho/1$ and regarding Eq. (19) and Eq. (20), the standard deviation of $\Delta A$ and $\Delta B$ can be written as follows,

$$\sigma_A = S_C(I_{mean} + CI_{mean}), \sigma_B = S_C(I_{mean} - CI_{mean}).$$

According to the rule of the combined uncertainty from correlated input quantities, defined in GUM [32], the standard deviation $\sigma_C$ of the contrast noise in Eq. (17) can be expressed by Eq. (22) (refer to the soon coming PhD thesis [30] for details).

$$\sigma_C^2 = (\frac{\partial C}{\partial A})^2 \sigma_{I_A}^2 + (\frac{\partial C}{\partial B})^2 \sigma_{I_B}^2 + 2\frac{\partial C}{\partial A}\frac{\partial C}{\partial B}\sigma_A \sigma_B r(\Delta A, \Delta B) \quad (22)$$
$$= 1/2 S_C^2 (1-C^2)^2 (1-r(\Delta A, \Delta B)).$$

Here, we switch to the realistic case, where the speckle noise is not perfectly correlated in the different phase shifted images. The correlation coefficient between $\Delta A$ and $\Delta B$ is $r(\Delta A, \Delta B)$. For complete correlation ($r=1$), the errors $\Delta A$ and $\Delta B$ cancel each other. In practice, $r$ is dependent on the fringe frequency and in the range of 80% (Fig. 5). It has to be determined experimentally.

As Eq. (16) for specular surfaces, we obtain Eq. (23) for the height uncertainty at rough surfaces by combination of Eq. (14) and Eq. (22). Experiments show that the speckle noise of the contrast curve, at different depth positions, is highly correlated as well, so smoothing over $N$ sampling points will not much affect the height uncertainty. In other words, there is not much advantage to sample the contrast curve in many narrow steps. That is why there is no $N$-dependence of Eq. (23). This is an important result, as we aim for high speed.

$$\sigma_z = 0.2 S_C \frac{\lambda}{\sin^2 u} \frac{\frac{1}{C}-C}{\bar{\nu}(1-0.5\bar{\nu})} \sqrt{1-r(\Delta A, \Delta B)}. \quad (23)$$

Equation (23) displays quite interesting consequences: the height precision $\sigma_z$ depends on three terms which tell us about the physics of SIM at rough surfaces. The well-known term $S_C \lambda/\sin^2 u$ is generic for triangulation: Dorsch [33] found that the depth precision in a laser triangulation sensor cannot be better than $S_C \lambda/2\pi \sin^2 u$.

The term $(1/C-C)/[\bar{\nu}(1-0.5\bar{\nu})]$ reflects the influence of the MTF of the lens. The term diverges for high fringe frequencies and is approximately constant for medium frequencies.

Now to the term $[1-r(\Delta A, \Delta B)]^{1/2}$: Figure 5(a) displays the measured correlation coefficient $r(\Delta A, \Delta B)$, as a function of the fringe frequency. In the medium range, we find $r \sim 0.9$. It might be noteworthy that for $r=0.93$, Eq. (23) delivers the same precision for SIM than Dorsch [22] found for laser triangulation ($u$=triangulation angle):

$$\sigma_z = \frac{S_C \lambda}{2\pi \sin^2 u}. \quad (24)$$

To validate the theory stated in Eq. (23), SIM measurements of a roughness standard with $R_a$=0.8μm are performed (Fig. 5) with a micro objective 5x/0.15 (see curve 1) and compared with the theory (see curve 2). The correlation coefficient $r(\Delta A, \Delta B)$ was experimentally determined, as shown in Fig. 5(a). As mentioned, speckle noise is highly correlated (for lower frequencies, other noise sources become more important, the correlation drops in this region). To calculate the theoretical precision, $C_{max}$ (~MTF) and the speckle contrast $S_C \sim 0.6$ were measured and the values were put into Eq. (23). The comparison shows that theory and experiment display considerable agreement in the medium frequency domain $\bar{\nu} > 0.05$. As mentioned, the model is not appropriate for very low frequencies.

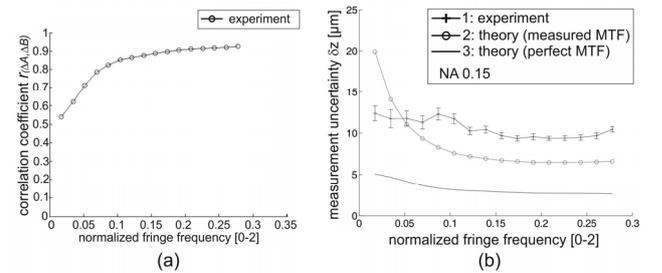

Fig. 5. Precision limit due to speckle noise: (a) experimentally determined correlation coefficient $r$ with 5x/0.15 micro objective. (b) Comparison of theoretical and experimental height uncertainty (again with 5x/0.15 micro objective, $\lambda$ = 0.5 μm and $C_{speckle}$=0.6.

As for specular surfaces, again there is no sharp minimum for the precsion, so there is a wide range of appropriate fringe frequencies. For high lateral resolution, of course, a high frequency is advantageous.

## 4. PRECISION, SPEED AND NUMBER OF DEPTH SCANNING STEPS

From the foregoing sections we can derive all necessary features of our system. According to the customer specifications, we start by selecting the NA of the lens and get the optimal fringe frequency as well as the achievable precision by using the introduced model. To estimate the best possible speed we have to calculate the minimal number $S_z$ of necessary z-scanning steps. The width FWHM of the contrast function is given by the Eq. (10) and Eq. (11), for smooth and rough objects, respectively. As mentioned above, the contrast curve needs a sampling distance $\Delta s \leq \Delta s_{max}$ = FWHM/2. With a given object depth $\Delta z$, the minimum number of necessary scanning steps is given by,

$$S_z = \frac{2\Delta z}{\text{FWHM}} + 1 + 4. \qquad (25)$$

The additive term comes up because we have to cover the edge of the object as well. For a rough object with 500 μm depth, and an NA=.15 lens, we need $S_z$=50 steps at the optimal fringe frequency $\bar{\nu}$ =0.5. As we use 4-phase shift, the number E of exposures is 200. With a given camera frame rate $F$, (and the given precision) the total measuring time cannot be shorter than $E/F$.

If the required precision is lower, it is appropriate to select a smaller aperture, and exploit a wider contrast curve. It is not appropriate to use a high aperture and a low fringe frequency, because the lateral resolution will be unnecessarily reduced.

Of course, the number of scanning steps can be increased over the minimal number $S_z$, in order to improve the precision via averaging. As already mentioned, this is not efficient for rough objects. For smooth objects it is helpful, because photon noise is uncorrelated in the different exposures.

## 5. SUMMARY AND CONCLUSION

The paper is originally motivated by the search for a 3D-metrology principle that has the potential for unprecedented high SBSP and for high precision. Moreover, high lateral resolution for both rough and specular surfaces is required. All these features are specifically necessary, e.g., for the inspection of electronic interconnects or for MEMS. The candidates for smooth and rough surface inspection are confocal microscopy (CM), white light interferometry (WLI) and Structured-Illumination Microscopy (SIM). The performance of WLI and CM is well known, but SIM is not yet well established for technical applications. This is why a physical model for the signal generation of SIM was developed. The model allows to estimate the fundamental limit of the precision, and its dependence from the system parameters such as aperture, fringe frequency and the number of depth sampling steps. With this knowledge it is possible to optimize SIM, for the best precision or for high speed.

We summarize the results by one example: With an NA=.15 lens, the optimal fringe frequency is in the range of 150 lp/mm. With a diffraction limited lens, the ultimate limit of the precision at rough objects is about 2 μm, in agreement with the well known limit of coherent triangulation. The precision limit on smooth surfaces is in the range of 200 nm, with standard video SNR ~100. With a very high aperture, NA=.85, the precision limit at smooth surfaces is about 5 nm. The comparison of our theoretical model with experiments displays a qualitative agreement, within the limits of experimental imperfections.

With the required precision and the object depth, given by the customer, our model enables the design of the optimal system parameters: the aperture, fringe frequency, number of scanning steps and maximum possible speed.

In part II of the paper, we will introduce a novel method, an apparatus and measurements displaying significantly faster scanning and a macroscopic field, without loss of precision and lateral resolution, suggesting SIM to be a candidate for a measuring system with very high space-bandwidth-speed product.

**FUNDING INFORMATION**

Deutsche Forschungsgemeinschaft (DFG) (HA 1319/13-1)

**ACKNOWLEDGMENT**

We would like to thank the Deutsche Forschungsgemeinschaft for the funding of this research as well as Markus Vogel for his support and the fruitful discussions.